\documentclass[final]{JHEP3} 

\JHEPspecialurl{http://jhep.sissa.it/JOURNAL/JHEP3.tar.gz}

\usepackage{epsfig,multicol,bbm}

\newcommand\fverb{\setbox\fverbbox=\hbox\bgroup\verb}
\newcommand\fverbdo{\egroup\medskip\noindent%
			\fbox{\unhbox\fverbbox}\ }
\newcommand\fverbit{\egroup\item[\fbox{\unhbox\fverbbox}]}
\newbox\fverbbox

\def\chioi{\tilde{\chi}^0_1}
\def\chioii{\tilde{\chi}^0_2}

\def\slr{\tilde{l}_R}
\def\sql{\tilde{q}_L}

\def\ra{\rightarrow}

\def\MT2{M_{T2}}

\def\m0{${0}$}

\title{A hybrid method for determining particle masses
at the Large Hadron Collider with fully identified cascade decays}

\author{Mihoko M. Nojiri, \\ Theory Group, KEK and the Graduate
	University for Advanced Study (SOUKENDAI), \\ 1-1 Oho,
	Tsukuba, 305-0801, Japan\\ and \\ Institute for the Physics
	and Mathematics of the Universe (IPMU), University of Tokyo,
	\\ 5-1-5 Kashiwa-noHa, Kashiwa City, Chiba 277-8568, Japan \\
	E-mail: \email{nojiri@post.kek.jp}} \author{Giacomo
	Polesello,\\ INFN, Sezione di Pavia, Via Bassi 6, 27100 Pavia,
	Italy \\ E-mail: \email{giacomo.polesello@cern.ch}}
	\author{Daniel R. Tovey,\\ Department of Physics and
	Astronomy, \\ University of Sheffield, Hounsfield Road,
	Sheffield S3 7RH, UK\\ E-mail:\email{ daniel.tovey@cern.ch}}

\received{} 		
\accepted{}		

\preprint{KEK-TH-1207\\IPMU-07-0008}

\abstract{A new technique for improving the precision of measurements
of SUSY particle masses at the LHC is introduced. The technique
involves kinematic fitting of events with two fully identified decay
chains. We incorporate both event $E_T^{miss}$ constraints and
independent constraints provided by kinematic end-points in experiment
invariant mass distributions of SUSY decay products. Incorporation of
the event specific information maximises the information used in the
fit and is shown to reduce the mass measurement uncertainites by
$\sim$ 30\% compared to conventional fitting of experiment end-point
constraints for the SPS1a benchmark model.}

\keywords{SUSY, fit, end-point, hybrid}

\begin{document} 


\section{Introduction}\label{sec1}

Measurement of SUSY particle masses in R-parity conserving SUSY events
at hadron colliders such as the LHC is complicated by a number of
factors. First, imperfect detector hermeticity close to the beam-pipe
and our ignorance of the initial state parton momentum prevent
measurement of the final state momentum in the
$z$(beam)-direction. This removes one kinematic constraint on the
masses of particles generated in any given event. Second, the presence
of an invisible Lightest SUSY Particle (LSP) at the end of each decay
chain generates eight unknown four-momentum components in every
event. As a result the kinematics of SUSY events at hadron colliders
are typically highly under-constrained.

Several approaches to this general problem have been documented. Given
a sufficiently long decay chain, constraints on analytical
combinations of SUSY particle (`sparticle') masses can be obtained
from the positions of end-points in distributions of invariant masses
of combinations of visible SUSY-decay products (jets, leptons etc.)
\cite{Hinchliffe:1996iu}. The system of equations may be solved with a
numerical fit to obtain the individual masses, if enough constraints
are provided \cite{Hinchliffe:1996iu,Allanach:2000kt,Miller:2005zp}.

An alternative class of techniques which does not rely on the
end-point information is also available. These techniques vary the
assumed LSP momenta in a given event subject to event kinematic
constraints in order to construct variables sensitive to the masses of
sparticles present in the event. One such technique involves the
construction of the $M_{T2}$ variable
\cite{Lester:1999tx,Allanach:2000kt,Barr:2003rg}, which is usually
defined for events where the same decay chain appears in both `legs'
of the event. For given assumed LSP transverse momenta the maximum
value of the transverse masses of the two legs is calculated. This
quantity is then minimised by varying the LSP momenta subject to the
event $E_T^{miss}$ constraints to give the value of $M_{T2}$. The
end-point of the $M_{T2}$ distribution depends on the test LSP mass
$M_{test}$ and has recently been shown to display a kink structure
when $M_{test}$ is equal to the true LSP mass
\cite{Cho:2007qv,Gripaios:2007is,Barr:2007hy,Cho:2007dh}. This shows
that $E_T^{miss}$ constraints are important for obtaining the absolute
mass scale of SUSY particles.

Another, more involved technique of this kind is the {\it
mass relation method} \cite{Kawagoe:2004rz}. In this method the masses
of sparticles in the decay chain are calculated for a range of
LSP momenta. When the decay chain present in the event involves a
large number of SUSY states, the calculated sparticle masses are
constrained to lie on a hypersurface in the mass parameter space, and
the intersection of several such hypersurfaces from different events
determines the true sparticle masses. Alternatively the existence of
solutions to the system of mass-shell conditions provided by
under-constrained events can be used to constrain the individual
masses \cite{Cheng:2007xv}.

The techniques described above all seek to determine the masses of
particles appearing in kinematically under-constrained events by
combining information from multiple events, either explicitly, as in
the {\it mass-relation method}, or implicitly by obtaining constraints
from distributions of quantities derived from individual events. This
paper describes a simple extension of these techniques (the {\it
hybrid method}) in which information from multiple events is
combined with event information in a kinematic fit to fully
reconstruct events and hence constrain the masses of individual SUSY
particles.  In this paper, we especially emphasize a case where both
legs contain the same cascade decay chain, so that events are more
prominent over the background, and fewer unknown mass parameters are
required.  It is also possible to include $E_T^{miss}$ constraints for
this case, which provide independent information on the LSP mass. Full
reconstruction of SUSY particle kinematics with this technique could
also be useful for other purposes such as measuring the
spin-statistics of SUSY states.

Section~\ref{sec2} introduces the new technique while
Section~\ref{sec3} discusses a particular example of its application
and illustrates the improvement in mass measurement precision obtained
for a specific benchmark SUSY model. Section~\ref{sec4} concludes and
outlines directions for future work.

\section{Description of technique}\label{sec2}

To simplify the description, we first examine the case where a
specific decay chain appears in both legs of the event. Such
`symmetric' events will possess a considerably smaller branching ratio
than `single-leg' events containing only one such decay chain but the
additional decay products can lead to striking signatures which
strongly suppress the SM and SUSY backgrounds. Conversely, if such
events are observable then it is also probable that significant
numbers of single-leg events can also be observed. Both classes of
event are used by the technique described here.

Our strategy is as follows. We start with constraints on sparticle
masses obtained from kinematic end-points in distributions of
invariant mass combinations obtained from single-leg events, as
described in Section~\ref{sec1}. Because one of each of these
distributions is obtained from the experimental data-set considered we
call these constraints `experiment-wise'. In addition to such
constraints, we possess `event-wise' constraints which are obtained
from mass-shell conditions provided by the event visible decay
products and LSP momenta, where the latter are further constrained by
the measured event $E_T^{miss}$ components. A kinematic fit to both
the experiment-wise and event-wise constraints can possess fewer
degrees-of-freedom than the simple experiment-wise end-point fit,
potentially improving the sparticle mass precision.

In practice event visible momenta and $E_T^{miss}$ values may be
measured with limited precision causing each individual kinematic fit
to generate mass values which potentially deviate more from the true
masses than the equivalent quantities obtained from the experiment-wise
end-point fit. Nevertheless because the data-set provides us with a
number of uncorrelated candidate events, the means of all the fitted
event mass values can indeed improve the mass measurement
precision. The situation is analagous to measuring the mass of a
resonance decaying to visible decay products - the mass value obtained
from one event may be relatively inaccurate but the uncertainty is
reduced by considering the mean mass obtained from all events. 

It is instructive to consider at this point the roles played by the
different constraints in the event-wise fits in this simplified case
of symmetric events. Each leg of the event contributes four unknown
LSP four-momentum components to the degrees-of-freedom of the event,
while for a decay chain consisting of $n$ steps each leg contributes
$n$ mass-shell conditions. Let us define the number of
degrees-of-freedom of the experiment-wise end-point fit to be $d_{end}
= n_{mass} - c_{end}$, where $n_{mass}(=n)$ is the number of sparticle
masses and $c_{end}$ is the number of uncorrelated end-point
constraints. In this case the {\it extra} number of degrees-of-freedom
generated by including the event-wise information is $d_{evt} =
n_{mom} - c_{evt}$, where $n_{mom}$ is the number of unknown LSP
four-momentum components and $c_{evt}$ is the number of event-wise
constraints. The condition for inclusion of the event-wise information
to potentially improve the mass precision is $d_{evt}<0$.

For symmetric events $d_{evt}$ is given by $4-n$ for one leg and
respectively $6-2n$ or $8-2n$ for two legs with or without
$E_T^{miss}$ constraints applied. Consequently if $n>4$ even
considering one leg alone in conjunction with the end-point
constraints potentially improves the mass measurement precisions,
while if $n>3$ improved precisions can potentially be obtained by
using both legs with $E_T^{miss}$ constraints.

It should be noted that given an unambiguous assignment of visible
final states to decay chain steps, the technique described here could
be extended to non-symmetric events where some/all of the SUSY states
appear in both legs of the event. In this case, for $n_1$ and $n_2$
steps in legs 1 and 2, $d_{evt}$ is given by $6-(n_1+n_2)$ or
$8-(n_1+n_2)$ when including or excluding the event-wise $E_T^{miss}$
constraints. Consequently if $n_1+n_2>6$ improved mass measurement
precisions can be obtained by using event-wise fits with $E_T^{miss}$
constraints applied.

\section{Example: $n=4$ step symmetric decay chain}\label{sec3}

\subsection{General discussion}\label{subsec3.1}

Let us now consider an example of mass measurement using symmetric
events, for the specific case of $n=4$. The decay chain present in both legs is
\begin{equation}
  \delta \rightarrow \gamma  c \rightarrow \beta  bc \rightarrow \alpha abc,
\end{equation}
where Greek letters denote SUSY states, Roman letters denote visible
SM decay products and $\alpha$ is the LSP. Denoting the two legs of
the event with subscripts 1 and 2, the eight mass-shell conditions are:
\begin{eqnarray}
       (p(a_1)+p(b_1)+p(c_1)+p(\alpha_1))^2 & = & (p(a_2)+p(b_2)+p(c_2)+p(\alpha_2))^2 = m_{\delta}^2,\cr
       (p(a_1)+p(b_1)+p(\alpha_1))^2 & = & (p(a_2)+p(b_2)+p(\alpha_2))^2 = m_{\gamma}^2,\cr
       (p(a_1)+p(\alpha_1))^2 & = & (p(a_2)+p(\alpha_2))^2 = m_{\beta}^2,\cr
       (p(\alpha_1))^2 & = & (p(\alpha_2))^2 = m_{\alpha}^2.
\label{eqn1}
\end{eqnarray}
To these constraints should be added the two constraints provided by
the event $E_T^{miss}$ components:
\begin{eqnarray}
        p_x(\alpha_1)+p_x(\alpha_2) & = & E_x^{miss}, \cr
        p_y(\alpha_1)+p_y(\alpha_2) & = & E_y^{miss},
\label{eqn2}
\end{eqnarray}
thus giving ten event-wise constraints in total
(i.e. $c_{evt}=10$). The number of unknown parameters appearing in
these constraints is twelve -- eight LSP four-momentum components
together with four unknown sparticle masses. With the definitions from
Section~\ref{sec2} the number of extra unknown parameters present in
the event-wise fit relative to the end-point fit is therefore
$n_{mom}=8$. Consequently $d_{evt}= -2$ indicating a potential gain
over the end-point fit. In practice the four mass-shell conditions for
each leg may be solved analytically to give a locally invertible map
from the four sparticle masses to the four components of the LSP four
momentum. In this case $c_{evt}=2$ and $n_{mom}=0$, however $d_{evt}$
is clearly unchanged.

It is also interesting to consider the above example when $E_T^{miss}$
constraints are not included in the event-wise fit. In this case
$n_{mom}=8$ and $c_{evt}=8$ (or equivalently $n_{mom}=0$ and
$c_{evt}=0$ if solving the mass-shell conditions) and hence
$d_{evt}=0$. Each leg of the event is independent and the event-wise
constraints just map sparticle masses to LSP four momenta. The
event-wise fit is therefore equivalent to solving the kinematic
end-point mass constraints and no gain in mass precision is
obtained. The situation changes when the $E_T^{miss}$ constraints are
applied because in this case the kinematics of the two legs are
connected.

Note that an alternative, related, approach to this $n=4$ problem can
also be taken. Namely, one merely utilises the equivalence of the
fitted masses in the two legs of the event. In this case the
constraints in Eqns.~\ref{eqn1} and~\ref{eqn2} reduce in number to six
(since the masses appearing in Eqn.~\ref{eqn1} are {\it a priori}
unconstrained) and hence two degrees-of-freedom remain
($d_{evt}=2$). Thus each event is under-constrained. The LSP
four-momenta which satisfy the constraints for a given event define a
two-dimensional surface in the four-dimensional $m_{\alpha}$,
$m_{\beta}$, $m_{\gamma}$, $m_{\delta}$ space. If a few events of this
kind can be found, it is possible to solve for the masses, which are
given by the coordinates of the point of intersection of all the event
2D surfaces in the 4D space. This approach is similar to the mass
relation method described in the introduction and will be studied in
full in a future paper.

\subsection{Concrete example: $\sql$ decays at the SPS1a mSUGRA benchmark point}\label{subsec3.2}

Let us now examine in detail a concrete realistion of the decay
chain discussed above. At mSUGRA point SPS1a there is a significant
branching ratio for the decay chain 
\begin{equation}
\sql \ra \chioii q \ra \slr lq \ra \chioi llq.
\label{eqn4}
\end{equation} 
This chain provides 5 kinematic end-point mass constraints from
invariant mass combinations of jets and leptons
\cite{Weiglein:2004hn}:

\begin{itemize}

\item $m(ll)^{max}$ = 77.08 $\pm$ 0.08(scale) $\pm$ 0.05(stat) GeV
\item $m(llq)^{max}$ = 431.1 $\pm$ 4.3(scale) $\pm$ 2.4(stat) GeV
\item $m(llq)^{min}$ = 203.0 $\pm$ 2.0(scale) $\pm$ 2.8(stat) GeV
\item $m(lq)_{hi}^{max}$ = 380.3 $\pm$ 3.8(scale) $\pm$ 1.8(stat) GeV
\item $m(lq)_{lo}^{max}$ = 302.1 $\pm$ 3.0(scale) $\pm$ 1.5(stat) GeV

\end{itemize}

For this study unbiased samples equivalent to 100~fb$^{-1}$ (one Monte Carlo
`experiment') of SPS1a signal events and $t\bar{t}$ background events
were generated with {\tt HERWIG 6.4}
\cite{Corcella:2000bw,Moretti:2002eu} and passed to a generic LHC
detector simulation \cite{RichterWas:2002ch}. A lepton reconstruction
efficiency of 90\% was assumed.

Events were selected in which the above decay chain appears in both legs of
the event with the following requirements:

\begin{itemize}

\item $N_{jet}$ $\geq$ 2, with $p_T(j2)$ $>$ 100 GeV,
\item $M_{eff2} = E_T^{miss} + p_T(j1) + p_T(j2)$ $>$ 100 GeV,
\item $E_T^{miss}$ $>$ max(100 GeV,$0.2M_{eff2}$),
\item $N_{lep}$ $=$ 4, where $lep = e/\mu$(isolated) and $p_T(l4)$ $>$ 6 GeV,
\item 2 Opposite Sign Same Flavour (OSSF) lepton pairs. If the pairs
are of different flavour both pairs must have $m(ll)<m(ll)^{max}$. If
both pairs are of the same flavour then one and only one of the two
possible pairings must give two $m(ll)$ values which are both less
than $m(ll)^{max}$. These pairings allocate the leptons to each leg of
the event.
\item One and only one possible pairing of the two leading jets with
the two OSSF lepton pairs must give two $m(llq)$ values less than
$m(llq)^{max}$. These pairings allocate the jets to each leg of the
event.
\item For each inferred leg of the event the maximum(minimum) of the
two $m(lq)$ values must be less than $m(lq)_{hi(lo)}^{max}$. This
ordering allocates the leptons to the $near$ and $far$
\cite{Allanach:2000kt} positions in the decay chain.

\end{itemize}

The requirement of 4-leptons in two OSSF pairs and two high-$p_T$ jets
consistent with kinematic end-points, together with large
$E_T^{miss}$, is effective at removing the majority of SM and SUSY
backgrounds (see below).

Each selected event was fitted with MINUIT \cite{James:1975dr}. Free
parameters were taken to be the four masses appearing in the decay
chain: $m(\sql)$, $m(\chioii)$, $m(\slr)$ and $m(\chioi)$. In the
spirit of the discussion in Section~\ref{subsec3.1} the mass-shell
conditions and measured momenta of the visible decay products for each
leg were solved to determine the LSP four-momenta, giving two
solutions for each leg. The $\chi^2$ minimisation function was defined
by:
\begin{eqnarray}
\label{eqn3}
\chi^2 &=&   \left(\frac{m(ll)^{max}_{evt}-m(ll)^{max}_{expt}}{\sigma_{m(ll)^{max}}}\right)^2 \nonumber \\
       &+& \left(\frac{m(llq)^{max}_{evt}-m(llq)^{max}_{expt}}{\sigma_{m(llq)^{max}}}\right)^2
       + \left(\frac{m(llq)^{min}_{evt}-m(llq)^{min}_{expt}}{\sigma_{m(llq)^{min}}}\right)^2 \nonumber\\
       &+& \left(\frac{m(lq)_{hi;evt}^{max}-m(lq)_{hi;expt}^{max}}{\sigma_{m(lq)_{hi}^{max}}}\right)^2
       + \left(\frac{m(lq)_{lo;evt}^{max}-m(lq)_{lo;expt}^{max}}{\sigma_{m(lq)_{lo}^{max}}}\right)^2 \nonumber\\
       &+& \left(\frac{p_x(\tilde{\chi}_{1}^0(1))+p_x(\tilde{\chi}_{1}^0(2))-E_x^{miss}}{\sigma_{E_x^{miss}}}\right)^2
       + \left(\frac{p_y(\tilde{\chi}_1^0(1))+p_y(\tilde{\chi}_{1}^0(2))-E_y^{miss}}{\sigma_{E_y^{miss}}}\right)^2,
\end{eqnarray}
where $evt$ denotes an expected end-point value derived from the
masses in the event-wise fit with the formulae of
Ref.~\cite{Allanach:2000kt}, and $expt$ denotes a `measured'
experiment-wise end-point value. The uncertainties $\sigma$ in these
`measured' endpoints were those quoted above. The uncertainties on the
measurements of the $x$ and $y$ components of $E_T^{miss}$,
$\sigma_{E_x^{miss}}$ and $\sigma_{E_y^{miss}}$, were given by
$0.5\sqrt{E_T^{sum}}$ where $E_T^{sum}$ is the scalar sum of jet $p_T$
of the event. This function incorporating both event-wise $E_T^{miss}$
constraints and experiment-wise end-point constraints was evaluated
for each of the four pairs of $\chioi$ momentum solutions obtained
from solving the leg mass-shell conditions. Fitted masses were
obtained when $\chi^2$ was minimsed for the event. Fitted masses were
used in the subsequent analysis only if MINUIT judged the fit to have
converged and $\chi^2_{min} < 35.0$.

Following application of the selection cuts described above and the
requirements of fit convergence and low fit $\chi^2_{min}$ 38 SUSY
`signal' events with the above decay chain appearing in both legs were
observed. 4 SUSY background events were observed, consisting of the
above decay chain in both legs but with one or two leptonically
decaying staus produced in the decays of the $\chioii$'s. No
$t\bar{t}$ background events were observed in 100~fb$^{-1}$ equivalent
data. More SM background events may be expected in a real experiment,
given that effects such as charge and lepton mis-identification are
not included in the fast detector simulation. The use of full {\tt
GEANT} detector simulation is required to model correctly these
effects, which is beyond the scope of this paper, however given the
hard kinematic selection cuts it is reasonable to assume that they are
at least smaller than the negligible $t\bar{t}$ background considered above.

\FIGURE[ht]{
\epsfig{file=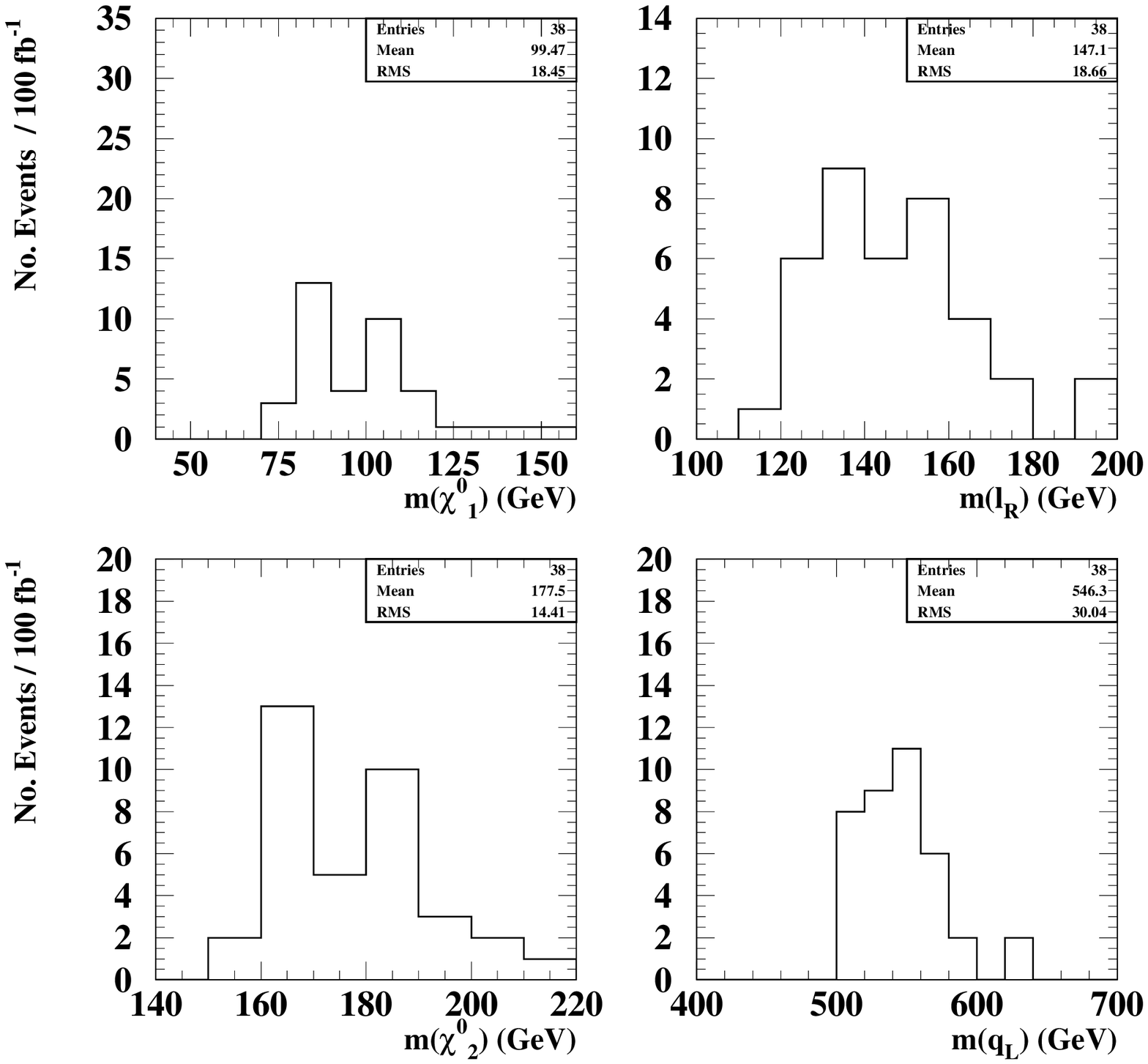,height=4.0in}
\caption{\label{fig1} Distributions of sparticle masses obtained from
event-wise fits, for one MC experiment. Each entry is obtained by
minimising the $\chi^2$ function shown in Eqn.~\ref{eqn3}. } }

\FIGURE[ht]{
\epsfig{file=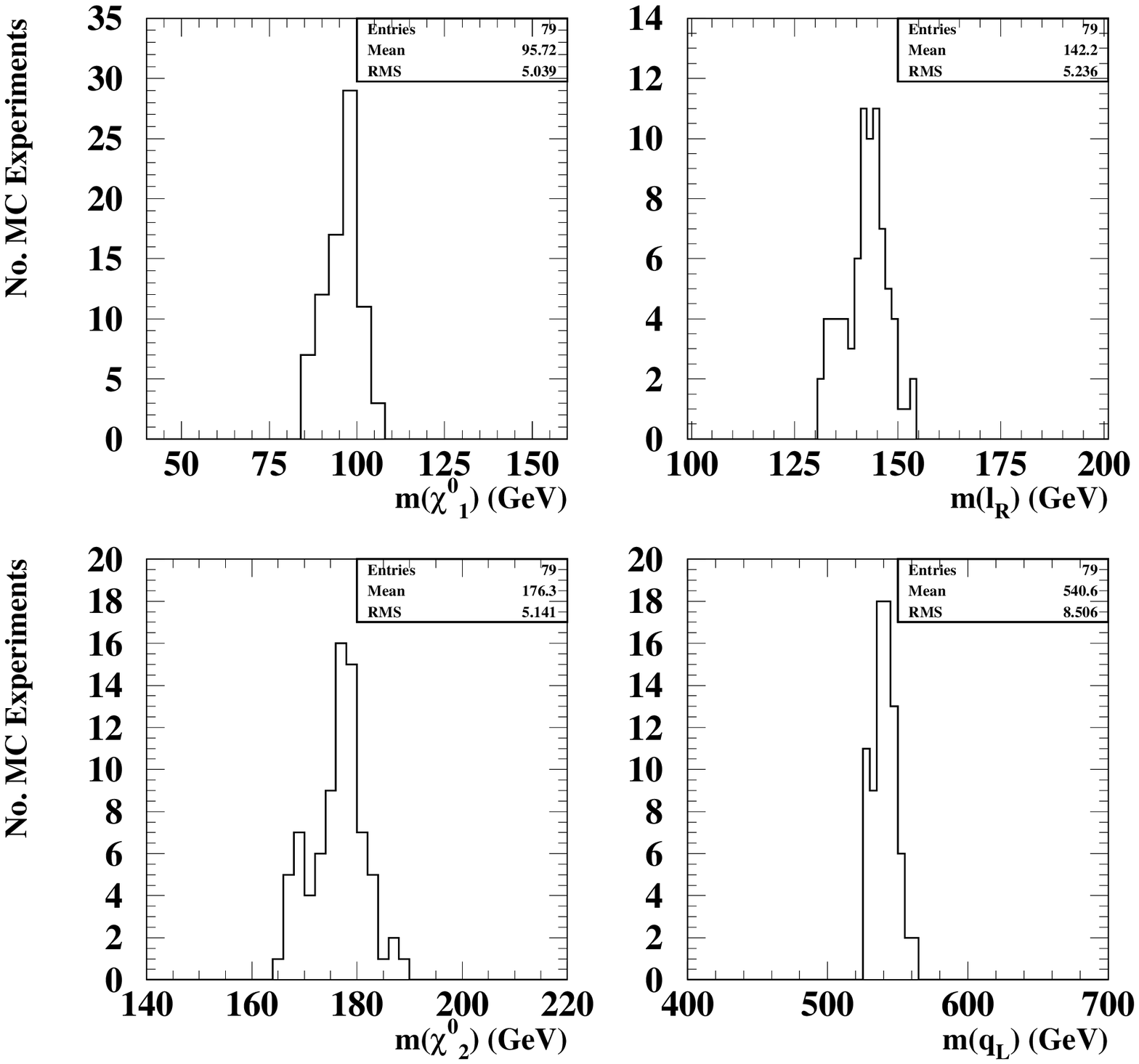,height=4.0in}
\caption{\label{fig2} Likelihood distributions of sparticle masses
obtained from 100 MC experiments. Each entry is the mean of an
experiment-wise mass histogram such as those in Fig.~\ref{fig1}.} }

Each event-wise fit generated one set of values for the sparticle
masses, namely those values which minimise the broad $\chi^2$ function
in Eqn.~\ref{eqn3}. The distributions of these values for one Monte
Carlo experiment are shown in Fig.~\ref{fig1}. The widths of the
distributions are greater than those which would be obtained were the
$E_T^{miss}$ constraints excluded from the fits -- in that case each
event-wise fit becomes equivalent to the experiment-wise end-point fit
as explained in Section~\ref{subsec3.1} and each distribution becomes
a delta-function located at the equivalent mass value. The key point
is that despite the fact that the mass distributions in
Fig.~\ref{fig1} are broader when $E_T^{miss}$ constraints are used,
the means of the distributions measure the sparticle masses more
accurately. This may seem counter-intuitive, however the reason is
clear. The $E_T^{miss}$ components for each event are measured with
only limited precision, causing a potential shift in the minimum of
the event $\chi^2$ function away from the true sparticle masses. There
is no way to correct for this effect on an event-wise basis. By
correctly parameterising the $E_T^{miss}$ measurement precision in
Eqn.~\ref{eqn3} however we ensure that such shifts are on average
unbiased when considering all events in the experiment and hence the
means of the mass distributions compensate for this resolution effect,
while also making use of the reduction in degrees-of-freedom provided
by the $E_T^{miss}$ constraints. This compensation for resolution
effects at the level of the experiment rather than individual events
is also implicit in the end-point method where gaussian-smeared end-point
functions are used when fitting the end-points in order to compensate
for experimental jet and lepton energy resolutions.

In order to demonstrate the performance of the technique and judge the
uncertainties in the measurements the above procedure was repeated for
100 Monte Carlo experiments. For each experiment, kinematic end-point
positions were sampled from gaussians with means and sigmas given by
the means and uncertainties listed above. The five sampled end-point
positions for each experiment were solved simultaneously with a MINUIT
fit to give initial mass values for input to the MINUIT event-wise
kinematic fits. For each experiment relative jet(lepton) energy scale
values were sampled from gaussians of width 1\%(0.1\%) reflecting
likely ultimate energy scale uncertainties at the LHC. Each experiment
generated a set of sparticle mass histograms similar to those shown in
Fig.~\ref{fig1}. The means of these histograms for the 100 MC
experiments were then used to construct likelihood histograms for the
masses, shown in Fig.~\ref{fig2}. The standard deviations of these
histograms were taken to provide the uncertainties on the sparticle
mass measurements.

Unbiased MC data equivalent to only one 100~fb$^{-1}$ experiment were
available for this study. For this reason the same events were used
for each MC experiment, with just the end-point values and jet/lepton
energy scales varying. The additional uncertainties in the final mass
values expected from varying event samples were estimated from the
mean statistical uncertainties in the mean experiment mass values as
extracted from the event-wise distributions such as those shown in
Fig.~\ref{fig1}. We evaluated the experiment-by-experiment spread due
to varying event samples as $\sigma/\sqrt{n}$, where $\sigma$ is the
RMS of the event-wise distributions as shown in Fig.~\ref{fig1}, and
$n$ is the number of entries in each plot. These additional
contributions were added in quadrature to the uncertainties obtained
from the study. This approximation was checked with a second sample of
SPS1a events equivalent to 100 different MC experiments, biased to
force gluinos to decay to $\sql$, $\tilde{b}$ or $\tilde{t}$, $\sql$
to decay to $\chioii$ and $\chioii$ to decay to $\tilde{e}$ or
$\tilde{\mu}$. Excluding the SUSY background from this sample
was estimated to bias the mean mass values by $<$ 1\% and increase the
uncertainties by $\lesssim$5\%. The uncertainties increase when the
SUSY background is excluded because the effect of the decrease in
event statistics outweighs the reduction in bias caused by excluding
the primarily $\tilde{\tau}$ background events, which have similar
kinematics to the $\tilde{e}$ and $\tilde{\mu}$ signal events.

\TABLE[ht]{\small%
\begin{tabular}{|c|c|c|c|c|c|c|c|}
\hline
State &Input &\multicolumn{2}{|c|}{End-Point Fit} &\multicolumn{2}{|c|}{Hybrid Method, $E_T^{miss}$}&\multicolumn{2}{|c|}{Hybrid Method, no $E_T^{miss}$}\\
\cline{3-8}
 & &Mean &Error &Mean &Error &Mean &Error \\
\hline
$\chioi$ &96.05     &96.5 &8.0 &95.8(92.2)   &5.3(5.5) &97.7(96.9)   &7.6(8.0) \\ 
$\slr$ &142.97      &143.3 &7.9 &142.2(138.7) &5.4(5.6) &144.5(143.8) &7.8(8.1) \\
$\chioii$ &176.81   &177.2 &7.7 &176.4(172.8) &5.3(5.4) &178.4(177.6) &7.6(7.9) \\
$\sql$ &537.2--543.0&540.4 &12.6 &540.7(534.8) &8.5(8.7) &542.9(541.4) &12.2(12.7) \\
\hline
\end{tabular}
\caption{Summary of mass measurement precisions for SPS1a
states. Column 2 lists masses used in the {\tt HERWIG} generator,
Columns 3 and 4 the fitted masses and uncertainties obtained from the
conventional fit to kinematic end-points, Columns 5 and 6 the
equivalent values obtained with the new technique and Columns 7 and 8
the equivalent values obtained with the new technique excluding
$E_T^{miss}$ constraints. Figures in parentheses are those obtained
with the biased sample of non-repeated events. All masses are in
GeV. The quoted mass range for $\sql$ excludes $\tilde{b}$ squarks,
which are produced less readily than the light squarks. \label{tab1}}}

The results of this study are summarised in Table~\ref{tab1}. For
comparison purposes the analysis was initially carried out with the
$E_T^{miss}$ constraints removed from the $\chi^2$ function. The
measurement precisions are consistent with those obtained from the
conventional end-point fitting method, as expected following the
reasoning outlined in Section~\ref{subsec3.1}. The analysis was then
repeated including the $E_T^{miss}$ constraints, giving an overall
improvement in sparticle mass precisions $\sim$ 30\% for all four
masses considered. A similar improvement was found when using the
biased sample of non-repeated events for different experiments. The
measurement of mass differences is also improved, with for instance
$m(\sql)-m(\chioi)$ being measured with a precision of 4 GeV
comparable with the natural widths of the light squarks ($\sim$ 5 GeV
at SPS1a) and their mass differences ($\sim$ 6 GeV). Further
improvement in mass measurement precision therefore probably requires
that such effects be taken into account, for instance in the
definition of the $\chi^2$ function in Eqn.~\ref{eqn3}.

\section{Conclusions}\label{sec4}
A new technique for improving the precision of LHC mass measurements
has been outlined in which experiment-wise information, for instance
invariant mass end-point constraints, are combined with event-wise
kinematic information such as $E_T^{miss}$ constraints and measured
four-momenta of visible decay products. For the SPS1a model considered
here the mass measurement precision was shown to improve by $\sim$
30\% for 100~fb$^{-1}$ of data. SUSY models with larger branching
ratios for symmetric events containing the necessary decay chain, or
with more indistinct or poorly-measured kinematic end-points, might be
expected to benefit still further from application of the
technique. The technique could be extended to non-symmetric events
where some/all of the SUSY states appear in both legs of the event --
this strategy for measuring further SUSY masses will be examined in
more detail in a future paper. Full reconstruction of SUSY particle
kinematics with this technique could also potentially be useful for
other purposes such as measuring the spin-statistics of SUSY states.

\section*{Acknowledgements}
The authors wish to thank the organisers of the Les Houches 2007 {\it
Physics at TeV-Scale Colliders} workshop, where the work described in
this paper was first conceived. DRT wishes to acknowledge STFC for
support.

\end{document}